\batchmode
\makeatletter
\def\input@path{{D:/Seafile/AMEM/Publikationen/2D/Submission/}}
\makeatother
\documentclass[english,aps, superscriptaddress, twocolumn]{revtex4-1}
\usepackage[T1]{fontenc}
\usepackage[latin9]{inputenc}
\usepackage{color}
\usepackage{babel}
\usepackage{units}
\usepackage{textcomp}
\usepackage{amsmath}
\usepackage{amssymb}
\usepackage{graphicx}
\usepackage{esint}
\usepackage{siunitx}
\usepackage[unicode=true,pdfusetitle,
 bookmarks=true,bookmarksnumbered=false,bookmarksopen=false,
 breaklinks=false,pdfborder={0 0 1},backref=false,colorlinks=false]
 {hyperref}

\makeatletter

\providecommand{\tabularnewline}{\\}

\usepackage{graphicx}

\makeatother

\begin{document}

\title{Autocorrected Off-axis Holography of 2D Materials }

\author{Felix Kern}

\affiliation{Institute for Solid State Research, IFW Dresden, Helmholtzstr. 20,
01069 Dresden, Germany}

\author{Martin Linck}

\affiliation{Corrected Electron Optical Systems GmbH, Englerstr. 28, D-69126 Heidelberg,
Germany}

\author{Daniel Wolf}

\affiliation{Institute for Solid State Research, IFW Dresden, Helmholtzstr. 20,
01069 Dresden, Germany}

\author{Nasim Alem}

\affiliation{Department of Materials Science and Engineering, Pennsylvania State
University,N-210 Millennium Science Complex University Park, PA 16802,
United States}

\author{Himani Arora}

\affiliation{Helmholtz-Zentrum Dresden-Rossendorf, Dresden 01328, Germany}

\author{Sibylle Gemming}

\affiliation{Helmholtz-Zentrum Dresden-Rossendorf, Dresden 01328, Germany}

\author{Artur Erbe}

\affiliation{Helmholtz-Zentrum Dresden-Rossendorf, Dresden 01328, Germany}

\author{Alex Zettl}

\affiliation{Department of Physics, University of California Berkeley, 366 LeConte
Hall MC 7300 Berkeley, CA, California 94720-7300, USA}

\author{Bernd B\"uchner}

\affiliation{Institute for Solid State Research, IFW Dresden, Helmholtzstr. 20,
01069 Dresden, Germany}

\author{Axel Lubk}

\affiliation{Institute for Solid State Research, IFW Dresden, Helmholtzstr. 20,
01069 Dresden, Germany}
\begin{abstract}
The reduced dimensionality in two-dimensional materials leads a wealth
of unusual properties, which are currently explored for both fundamental
and applied sciences. In order to study the crystal structure, edge
states, the formation of defects and grain boundaries, or the impact
of adsorbates, high resolution microscopy techniques are indispensible.
Here we report on the development of an electron holography (EH) transmission
electron microscopy (TEM) technique, which facilitates high spatial
resolution by an automatic correction of geometric aberrations. Distinguished
features of EH beyond conventional TEM imaging are the gap-free spatial
information signal transfer and higher dose efficiency for certain
spatial frequency bands as well as direct access to the projected
electrostatic potential of the 2D material. We demonstrate these features
at the example of \textit{h}-BN, at which we measure the electrostatic
potential as a function of layer number down to the monolayer limit
and obtain evidence for a systematic increase of the potential at
the zig-zag edges. 
\end{abstract}
\maketitle

\section{Introduction}

The discovery of graphene and its intriguing properties more than
ten years ago \citep{Novoselov(2004),Geim(2007)} has sparked large
and ongoing research efforts into two-dimensional materials (2DMs).
The synthesis of novel 2DMs, comprising, e.g., 2D topological insulators,
2D magnets, or organic systems like 2D polymers, with single to few
layers thickness and high structural definition at the atomic / molecular
level is at the center of this field (e.g., \citep{Wang2012,Butler2013,Miro2014,Mannix2017}).
They exhibit a large range of physical properties triggered by the
reduced dimensionality in one direction such as quantum confinement
effects or weak dielectric screening from the environment, yielding
a significant enhancement of the Coulomb interaction \citep{Wang2012,LopezSanchez2013}.
Another interesting aspect is the formation of out-of-plane elastic
modulations, which stabilize the 2DM structure and modify its mechanical
properties \citep{Peierls1935,Meyer2007,Deng2016}. The missing third
dimension also enhances the proliferation and impact of defects, such
as point and line defects, grains or multilayers morphologies; which
inevitably occur upon synthesis and often govern the functionality
(e.g., reactivity, stability) of 2DMs in applications \citep{Lin2016}.

Therefore the development of microscopic characterization methods,
which allow to analyze the structure and electronic properties of
the 2DMs including the edges, defects and grain boundaries, is at
the center of the field. Transmission electron microscopy (TEM) has
been a cornerstone technique (others are scanning tunneling microscopy
STM and photo emission electron microscopy PEEM), offering high resolving
power and spectroscopic information. A breakthrough for TEM could
be achieved by employing chromatic aberration correction facilitating
high-spatial resolution at low-acceleration voltages \citep{Linck2016}.
Two central challenges required special attention and have been addressed
through various methodological developments of TEM techniques:

(A) 2D materials are typically more susceptible to various beam damage
mechanisms than their 3D counterparts \citep{Komsa2012,Lehnert2017,Zhang2018}.
That includes knock-on damage, radiolysis, and chemical etching. The
knock-on damage may be reduced by lowering the acceleration voltage
and hence the kinetic energy of the beam electron below the knock-on
threshold of the pertinent chemical bonds in the 2DM (e.g., 90 keV
for the C-C bond in graphene \citep{Meyer2012} and 40 keV for the
B-N bond in monolayer \emph{h}-BN \citep{Kotakoski2010}). Radiolysis
and etching follow more complicated reaction mechanisms \citep{Susi2019}.
Their magnitude might be reduced by lowering the temperature and optimizing
the acceleration voltage \citep{Meyer2012}.

(B) Most 2D materials belong to the family of weak scatterers (another
important member is biological matter, mainly consisting of C and
H atoms), which implies that they only (weakly) shift the phase of
the electron wave when traversing the sample, but don't modulate the
amplitude. Consequently, they are referred to as weak phase objects
(WPOs). This phase shift of the electron wave can not be measured
directly, due to the quantum mechanical phase detection problem. To
solve this problem one can employ phase plates, which enable the imaging
of the phase shift introduced by these materials as intensity contrast.
The use of either physical \citep{Boersch1947,Kanaya1958,Danev2001,Danev2014}
or electron optical phase plates \citep{Boersch1947,Matsumoto1996,Majorovits2007,Cambie2007},
however, comes with some merits and disadvantages. The former degrade
\citep{Danev2001,Schultheiss2006,Danev2014} during use and create
unwanted diffuse scattering and beam blocking \citep{Nagayama2008},
whereas the latter is typically constructed from materials that are
prone to charging. Notable exceptions are laser \citep{Schwartz2019}
and drift tube \citep{Cambie2007} phase plates, which are very demanding
construction- and implementation-wise. By far the most straight forward
method for transfer of phase contrast to intensity constrast, however,
is an additional defocus with respect to the object exit plane, which
has the negative side effect of introducing transfer gaps at low spatial
frequencies or an oscillating contrast transfer for large spatial
frequencies (see below). Introducing large defocii also results in
a reduction of the resolution due to the partial transversal coherence
of the electrons, which may be expressed by an exponential envelope
function in reciprocal space.

In the following we address the phase problem in weak scatterers (at
the example of 2DMs) by advancing off-axis electron holography, an
interferometric technique allowing to reconstruct the phase shift
of the electron wave over the whole spatial frequency band, up to
the information limit. These advantages have triggered a small number
of previous studies on WPOs, notably at biological materials \citep{Simon2004,Simon(2008)}
and 2DMs \citep{Ortolani2011,Cooper2014,Tavabi2015,Borghardt2017,Winkler2017,Winkler2018}.
However, a persisting problem remains in the defocus required for
visualizing the sample during TEM operation and other residual aberrations
such as astigmatism, which typically built up during acquisition \citep{Barthel(2010)}.
Their correction, however, is mandatory for an analysis of the acquired
phase in terms of physical quantities such as potentials, charge densities
and the atomic structure. In Winkler \textit{et al}. \citep{Winkler2017,Winkler2018}
this problem has been addressed by a model-based fitting approach
requiring a full model of the scattering potential and hence the 2DM
under investigation.

Here we follow a different approach requiring no or only minimal a-priori
knowledge of the sample, that is, stripping the recorded data from
any instrumental influences, notably aberrations and noise. The resulting data
may then be used to extract certain specimen properties in a second
step. This approach has the advantage of requiring no a-priori knowledge
about the specimen and a clear separation between instrumental and
specimen influences. One key idea is to exploit very general discrete
symmetries pertaining to the scattered electron wave function: First
of all, the weak scattering property induces an odd symmetry in the
object phases in Fourier space, which allows to correct for symmetric
aberrations. Second, a large class of 2DMs are centrosymmetric, introducing
an even symmetry in the Fourier object phases, which allows to correct
for antisymmetric aberrations. This approach is based on the original
work of Fu and Lichte \citep{Fu(1991)a}, who demonstrated how to
generically extract symmetric aberrations from holograms at the example
of amorphous carbon foils. Here, we go one step further and autocorrect
for the aberrations in the reconstructed wave of 2DMs, which greatly
facilitates the analysis of the phase in terms of physical data, i.e.,
projected potentials. In this regard, we follow D. Gabor's original
idea of holography as a means to aberration correction \citep{Gabor(1948)}.
The corrected data is then subjected to a principle component analysis
(PCA) denoising, which reveals the meaningful phase data at the atomic
scale and enables extracting the underlying projected potential. Last
but not least we compare that data to ab-initio density functional
calculations to analyze the measured potentials in terms of charge
(de)localizations.

The paper is organized as follows, we first recapitulate the imaging
principles of weak scatterers (phase objects) and off-axis holography.
From these, we derive how geometric aberrations can be determined
and automatically corrected \textit{a-posteriori} from an acquired
hologram without additional measurements. We elaborate on the noise
characteristics of the thereby obtained aberration-corrected image
phase and the spatial resolution of the determined aberrations (e.g.,
defocus due to out-of-plane modulations). We finally demonstrate the
feasibility of the approach at the example of \emph{h}-BN. Amongst
others we reconstruct the number of layers, the mean inner potential
(MIP) of individual layers, the structure of the monolayer as well
as the edges; and correlate this to material properties such as the
charge delocalization or the stability and electronic properties of
the edge states.

\section{Imaging Principles}

\subsection{Conventional imaging and off-axis holography of weak phase object}

\begin{figure}
\includegraphics[clip,width=1\columnwidth]{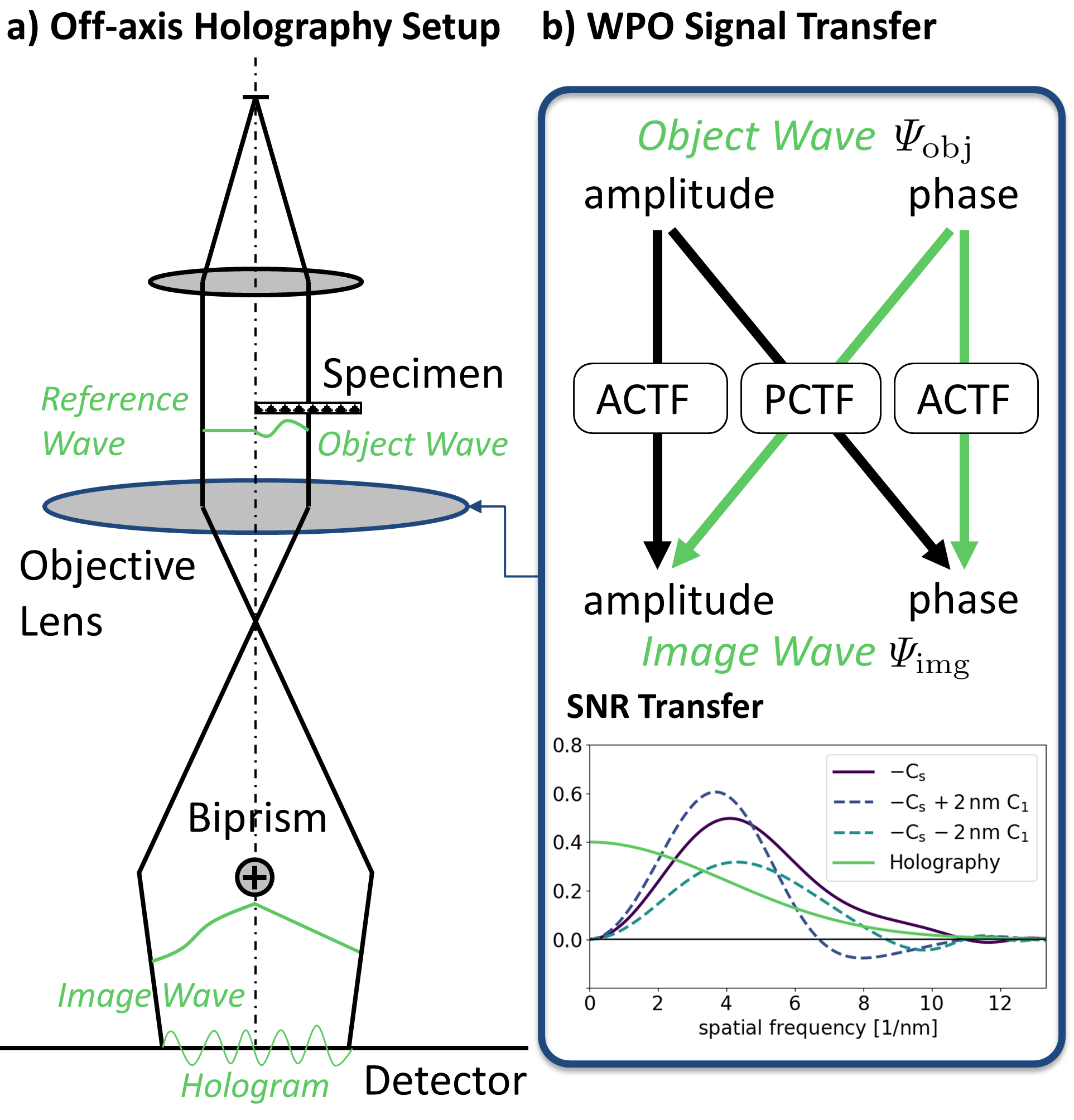}\caption{\label{fig:transfer-functions}(a) Off-axis electron holography (EH)
setup, (b) signal transfer for imaging a weak phase object (WPO) and
signal-to-noise (SNR) transfer functions for HRTEM in negative $C_{\mathrm{s}}$
conditions ($-C_{\mathrm{s}}$) and EH (c.f. Eq. \ref{eq:noise_ratio}).
In case of HRTEM, only the signal transfer from object phase to image
amplitude described by the phase contrast transfer function (PCTF)
contributes to the image. In case of EH, also the signal transfer
from object phase to image phase described by the amplitude contrast
transfer function (ACTF) contributes to the image. Additionally, the
corresponding transferred bands for HRTEM in $-C_{\mathrm{s}}$ conditions
with \textpm{} 2 nm defocus variation are also plotted. A defocus
drift of about 2 nm is commonly observed after about five minutes
in aberration-corrected TEM instruments \citep{Barthel(2010)}.}
\end{figure}

Sufficiently thin TEM specimens (with the critical thickness depending
on the atomic scattering potential $V$ of the chemical constituents)
behave as weak phase object (WPOs) in TEM. In good approximation they
just impose a small phase shift

\begin{equation}
\varphi_{\mathrm{obj}}\left(\boldsymbol{r}\right)=C_{\mathrm{E}}\intop_{-\nicefrac{t}{2}}^{+\nicefrac{t}{2}}V\left(\boldsymbol{r},z\right)\mathrm{d}z\label{eq:PGA}
\end{equation}
on the beam electrons' wave function $\varPsi_{\mathrm{obj}}$ leaving
the object of thickness $t$. Here $C_{\mathrm{E}}$ is an electron-energy-dependent
interaction constant ($0.01\,\nicefrac{\text{rad}}{V\,\mathrm{nm}}$
at 80 keV), $z$ the direction in which the electron beam transmits
the sample, and $\boldsymbol{r}$ the 2D position vector in the object
plane. Since WPOs do not modulate the amplitude $A$, $\varPsi_{\mathrm{obj}}$
can be approximated by
\begin{equation}
\varPsi_{\mathrm{obj}}\left(\boldsymbol{r}\right)\overset{\left(\text{WPO}\right)}{\approx}A_{\mathrm{obj}}\left(1+i\varphi_{\mathrm{obj}}\left(\boldsymbol{r}\right)\right).
\end{equation}
The dominant geometric aberrations of the objective lens are described
by a phase function $\chi\left(\boldsymbol{k}\right)$ acting on the
electron wave spectrum by a complex factor $e^{-i\chi\left(\boldsymbol{k}\right)}$,
the wave transfer function (WTF), in reciprocal space. Moreover, the
combination of (transversal and longitudinal) partial coherence and
geometric aberrations leads to an exponential damping of spatial frequencies
in the wave function described by a real envelope function $E\left(\boldsymbol{k}\right)$
in reciprocal space. The Fourier transform of the so-called image
wave function taking into account these modulations by the imaging
system (Fig. \ref{fig:transfer-functions}(a)) reads
\begin{eqnarray}
\mathrm{\tilde{\varPsi}_{\mathrm{img}}\left(\boldsymbol{k}\right)} & = & \tilde{\varPsi}_{\mathrm{obj}}\left(\boldsymbol{k}\right)e^{-i\chi\left(\boldsymbol{k}\right)}E\left(\boldsymbol{k}\right)\label{eq:img_Wave_k}\\
 & \overset{\left(\text{WPO}\right)}{\approx} & \mathrm{A_{obj}\left(\delta\left(\boldsymbol{k}\right)+{\color{white}e^{0^{0}}}\right.}\nonumber \\
 &  & \left.ie^{-i\left(\chi_{s}\left(\boldsymbol{k}\right)+\chi_{a}\left(\boldsymbol{k}\right)\right)}E\left(\boldsymbol{k}\right)\tilde{\varphi}_{\mathrm{obj}}\left(\boldsymbol{k}\right)\right)\,.
\end{eqnarray}
Here we separated the antisymmetric and symmetric aberrations, $\chi_{\mathrm{a}}\left(\boldsymbol{k}\right)$
and $\chi_{\mathrm{s}}\left(\boldsymbol{k}\right)$, respectively.
The corresponding conventional linear image intensity (neglecting
the term quadratic in $\varphi_{\mathrm{obj}}\left(\boldsymbol{r}\right)$
and additional smearing due to the detector) reads
\begin{eqnarray}
I\left(\boldsymbol{r}\right) & = & \left|\varPsi_{\mathrm{img}}\left(\boldsymbol{r}\right)\right|^{2}\label{eq:intens}\\
 & \approx & A^{2}+\nonumber \\
 &  & 2A^{2}\mathcal{F}^{-1}\left\{ \underset{\mathrm{PCTF}}{\underbrace{\sin\chi_{\mathrm{s}}\left(\boldsymbol{k}\right)E\left(\boldsymbol{k}\right)}}e^{-i\chi_{\mathrm{a}}\left(\boldsymbol{k}\right)}\right\} \ast\varphi_{\mathrm{obj}}\left(\boldsymbol{r}\right)\,.
\end{eqnarray}
Here we observe that only the so-called phase contrast transfer function
(PCTF) containing the symmetric aberrations produces visible contrast
by convolution ($\ast$) with the phase (Fig. \ref{fig:transfer-functions}(b)).
Several strategies have been developed to optimize the transfer over
certain spatial frequency bands. Most notably, in state-of-the-art
instruments equipped with hardware-aberration correctors, the spherical
aberration ($C_{\mathrm{s}}$) and defocus can be traded to produce
a positive contrast transfer for high resolution TEM (HRTEM) over
a band, ultimately limited by the incoherent chromatic envelope of
the instrument (also referred to as negative $C_{\mathrm{s}}$ imaging
conditions \citep{Lentzen2002}). The latter limitation could be largely
eliminated by employing chromatic aberration correctors \citep{Haider(2010)},
ultimately leading to an image spread limited resolution (due to Johnson
noise \citep{Uhlemann(2013)})
\begin{equation}
E\left(\boldsymbol{k}\right)=\exp\left(-2\pi^{2}\sigma_{\mathrm{i}}^{2}k^{2}\right)
\end{equation}
 in such instruments (Fig. \ref{fig:transfer-functions}(b)). Upon
inspection of the PCTF it becomes immediately clear that, if negative
$C_{\mathrm{s}}$ conditions are perfectly adjusted, it acts as a
band pass in HRTEM conditions, mainly suppressing small spatial frequencies
(-$C_{\mathrm{s}}$ in Fig. \ref{fig:transfer-functions}(b)). This
property complicates for instance the analysis of the 2DM's morphology
such as determining the layer number representing large scale spatial
structures (see, e.g. \citep{Alem2009}, for example on \emph{h}-BN).

While the transfer of large spatial frequencies may be only increased
through further improved electron optics and/or reconstructing multiple
images with varying imaging conditions (e.g., focal series, \citep{Coene1996},
beam tilt series\citep{Kirkland1995}, or Ptychography \citep{Maiden2009}),
both large and small spatial frequencies can be transferred simultaneously
by employing off-axis holography. Here, the object is inserted half-way
into the beam path, which automatically restricts the field of view
(FOV) to edges of the 2D material, whereas the other half-space is
occupied by the undisturbed reference wave. Both parts are brought
to superposition by employing an electrostatic M{\"o}llenstedt biprism,
forming a hologram in the image plane (Fig. \ref{fig:transfer-functions}(a)).
From the latter a complex wave function
\begin{align}
\tilde{\varPsi}_{\mathrm{hol}}\left(\boldsymbol{r}\right) & =\mathcal{F}^{-1}\left\{ \mu\left(\boldsymbol{k}\right)\tilde{\varPsi}_{\mathrm{img}}\left(\boldsymbol{k}\right)\right\} 
\end{align}
is reconstructed \citep{Lehmann2002}. Here, the contrast factor $\mu=\mu_{\mathrm{c}}\mathrm{MTF}$
takes into account the illumination degree of coherence, inelastic
scattering and instrumental instabilities (wrapped up in $\mu_{c}$) and the modulation transfer function (MTF)
of the detector.
Since the whole wave function is reconstructed only the incoherent
envelopes $E\left(\boldsymbol{k}\right)$ limit the transfer, in particular
there is no damping of small spatial frequencies by the PCTF. For
simplicity we will approximate the contrast factor $\mu\left(\boldsymbol{k}\right)$
with that of the hologram carrier frequency $\mu\left(\boldsymbol{k}_{\mathrm{c}}\right)$
in the following.

In addition, the linear reconstruction principle allows to compute
the noise transfer and hence the error (in terms of variance) pertaining
to the reconstructed phase from the noise transfer function of the
detector \citep{Lubk2012,Roder(2014)a}. If we use this result and
compare conventional HRTEM of WPOs with off-axis holography in terms
of signal-to-noise ratio (SNR) of the phase contrast for a particular
spatial frequency (see Appendix \ref{sec:Noise-Transfer} for a detailed
derivation), we obtain
\begin{equation}
\frac{\mathrm{SNR_{hol}}}{\mathrm{SNR_{conv}}}=\frac{\mu_{\mathrm{c}}\left(\boldsymbol{k}_{\mathrm{c}}\right)\sqrt{\text{DQE}\left(0\right)}}{\sin\chi_{\mathrm{s}}\sqrt{\text{DQE}\left(\boldsymbol{k}_{\mathrm{c}}\right)}}\,.\label{eq:noise_ratio}
\end{equation}
Here, DQE denotes a 2D generalization of the detection quantum
efficiency as detailed in Appendix \ref{sec:Noise-Transfer}. If this ratio
becomes larger than $1$, i.e. 
\begin{equation}
\frac{\mu_{\mathrm{c}}\left(\boldsymbol{k}_{\mathrm{c}}\right)\sqrt{\text{DQE}\left(0\right)}}{\mathrm{\sin\chi_{s}\sqrt{\text{DQE}\left(\boldsymbol{k}_{c}\right)}}}>1\,,
\end{equation}
off-axis holography is more dose-efficient than conventional phase
contrast in terms of retrievable information per dose. Noting that
a realistic value for the fringe contrast in high-resolution holograms
recorded at modern TEMs equipped with state-of-the-art detectors and
field-emission guns can reach several 10\% (in this work 30\%, see
below), this condition is met in a broad range of low to medium spatial
frequencies (see Fig. \ref{fig:transfer-functions}(b)) but not for
large spatial frequencies. Note, however, that the latter restriction
may be overcome by the use of novel direct counting detectors with
largely reduced detector DQEs \citep{Ruskin2013,McMullan2014}.

\subsection{A posteriori correction of residual aberrations\label{subsec:A-posteriori-correction}}

Following Fu and Lichte \citep{Fu(1991)a} the symmetric aberrations
can be readily extracted from the phases $\tilde{\varphi}_{\mathrm{img}}$
of the image wave function in reciprocal space (Eq. \ref{eq:img_Wave_k})
\begin{align}
\chi_{s} & (\boldsymbol{\boldsymbol{k}})=-\frac{1}{2}\left(\tilde{\varphi}_{\mathrm{img}}\left(\boldsymbol{\boldsymbol{k}}\right)+\tilde{\varphi}_{\mathrm{img}}\left(-\boldsymbol{\boldsymbol{k}}\right)\right)+\frac{\pi}{2}+\pi n\left(\boldsymbol{k}\right),\label{eq:chi_s}
\end{align}
which follows from the antisymmetry of Fourier phases of the original
WPO (see Appendix \ref{sec:WPO-Phase-Plate} for a detailed derivation).
Here the appearance of the integer $\left(n\epsilon\mathbb{N}\right)$
$\pi$-ambiguity stems from the $2\pi n$ ambiguity of the original
wrapped phases. The above relation is remarkable as it allows to compute
(and therefore correct) the symmetric part of the phase plate $\chi\left(\boldsymbol{\boldsymbol{k}}\right)$
(aberrations) without any \textit{a-priori} knowledge about the object
or the incoherent envelopes including the detector MTF. The only condition
for a successful practical application is that the object spectrums
SNR must be large enough to suppress error propagation of inevitable
reconstructed noise and other artifacts (e.g., Fresnel fringes).

A similar expression cannot be derived for the antisymmetric aberrations,
because they do not produce an amplitude variation from the WPO (see
Eq. (\ref{eq:intens})). Similar to the well-known Zemlin tableau
method \citep{Zemlin(1980)}, they can be determined from a tilt series
(where lower order symmetric aberrations are induced by higher order
antisymmetric ones), or additional symmetry criteria. In particular,
for the large class of centrosymmetric 2DMs we have $\tilde{\varphi}_{\mathrm{a}}=0,\,\pi$
and hence
\begin{equation}
\chi_{\mathrm{a}}(\boldsymbol{\boldsymbol{k}})=-\frac{1}{2}\left(\tilde{\varphi}_{\mathrm{img}}\left(\boldsymbol{\boldsymbol{k}}\right)-\tilde{\varphi}_{\mathrm{img}}\left(-\boldsymbol{\boldsymbol{k}}\right)\right)+\pi n\left(\boldsymbol{k}\right)\label{eq:chi_a}
\end{equation}
Here, $\left(n\epsilon\mathbb{N}\right)\times\pi$-ambiguity stems
from the $\pi$-phases of the object.

To finally correct for the aberrations from the holographically reconstructed
wave functions, we multiply its Fourier transform with the complex
conjugate of the WTF, i.e.
\begin{equation}
\tilde{\varPsi}_{\mathrm{obj}}\left(\boldsymbol{k}\right)E\left(\boldsymbol{k}\right)=\tilde{\varPsi}_{\mathrm{img}}\left(\boldsymbol{k}\right)e^{i\chi\left(\boldsymbol{k}\right)}\,.
\end{equation}
Note, however, that this involves a phase unwrapping procedure removing
the $\pi$-ambiguity in the phase plate, which can be challenging
in practice, depending on the spectrum of the object wave. This currently
limits the scope of the autocorrection scheme to pre-corrected imaging
conditions (e.g., using hardware corrected TEMs), where only small
residual aberrations and sufficiently small defoci are present, keeping
the phase range within $\pi$ over a large band.

The above considerations are strictly correct for the WPO only. In
praxis, this condition may be violated to some extend, e.g., when
employing low-acceleration voltages (resulting in higher phase shifts)
to reduce knock-on damage in a certain class of 2DMs. Note, however,
that the ``constant-amplitude'' criterion also applies to pure phase
objects and may be even slightly generalized to weak amplitude objects
by minimizing a penalty term for the amplitude variations, e.g.,
\begin{equation}
\chi\left(\boldsymbol{k}\right)=\arg\min\left\Vert \nabla\left|\tilde{\varPsi}_{\mathrm{img}}\left(\boldsymbol{k}\right)e^{i\chi\left(\boldsymbol{k}\right)}\right|\right\Vert \,.
\end{equation}
Lehmann \citep{Lehmann2000} and Ishizuka \textit{et al.} \citep{Ishizuka1994}
reported different approches to this aberration assesment via direct
amplitude variation minimization for WPOs. It is currently an open
question, whether and under which conditions this generalization yields
unique solutions \citep{Linck2010a}.

\section{Experimental}

\begin{figure}
\begin{centering}
\includegraphics[width=1\columnwidth]{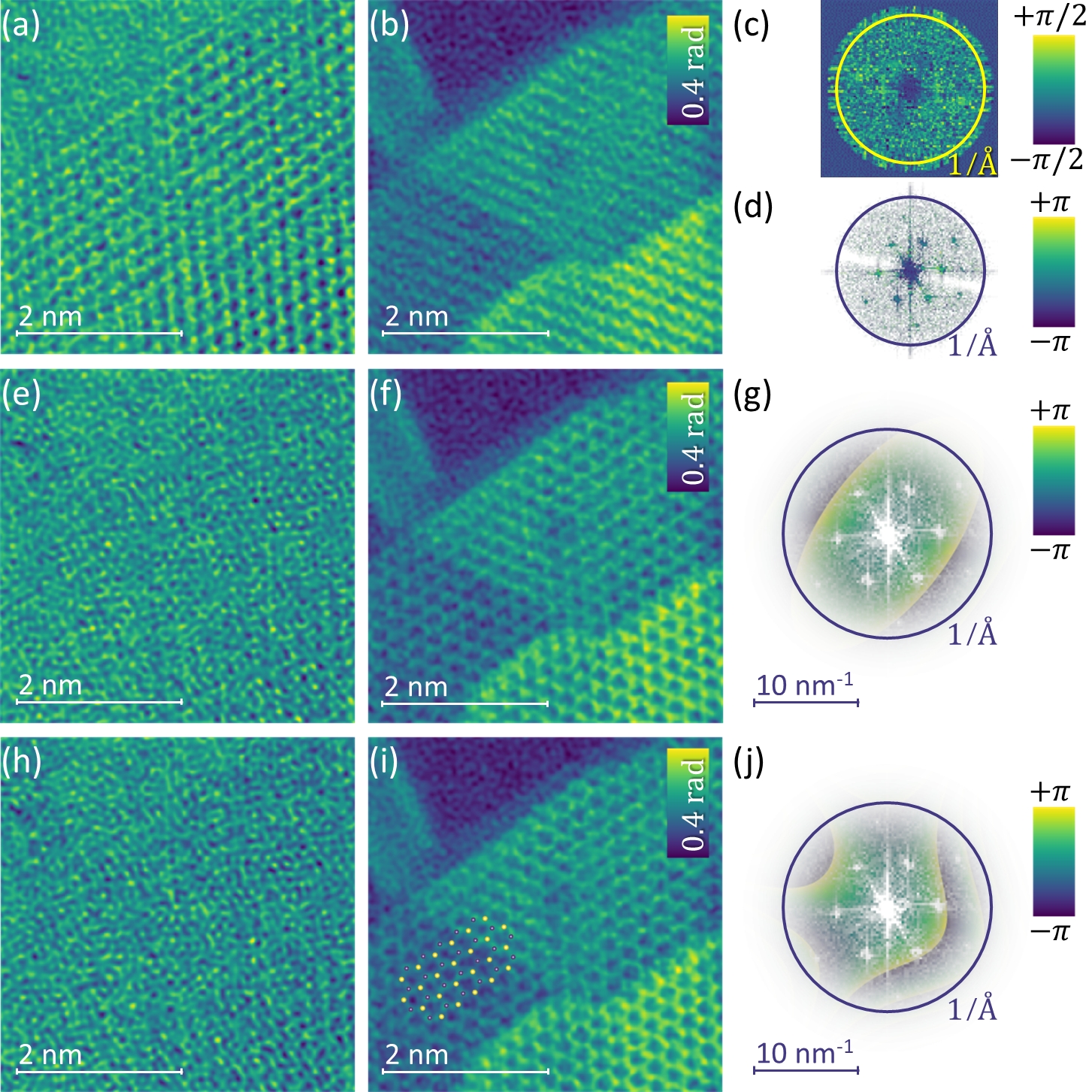}
\par\end{centering}
\caption{\label{fig:Autocorrection-procedure.}Aberration correction of \emph{h}-BN
image wave reconstructed by off-axis electron holography. (a) and
(b) show the as-reconstructed wave image in amplitude and phase. The
symmetric aberration phase plate (c) as determined from Eq. (\ref{eq:chi_s})
is not very evident, however, it becomes much clearer in (d) the color
coded representation of (c) overlayed with the Fourier spectrum of
the image wave (a,b). (e) and (f) depict amplitude and phase corrected
for symmetric aberrations using the continuous phase plate (g) fitted
from (c). (h) and (i) show amplitude and phase corrected also for
antisymmetric aberrations (coma and 3-fold astigmatism) obtained from
Eq. (\ref{eq:chi_a}) using the phase plate (j). The crystal structure
of a \emph{h}-BN monolayer is indicated in (i).}
\end{figure}

To validate the autocorrection theory we apply the above machinery
to a single to few atomic layer van-der-Waals 2DM, namely hexagonal
Boron Nitride \emph{(h}-BN). \emph{h}-BN has a crystal structure very
similar to that of graphene (see Fig. \ref{fig:Autocorrection-procedure.}(i)),
but possesses completely different electronic properties (notably
a large band gap, no Dirac points) \citep{Cassabois2016,Wang2017}.
The electron holograms have been recorded at a chromatic aberration
($C_{\mathrm{c}})$-corrected TEM instrument, the TEAM I at the Molecular
Foundry at the National Berkeley Lab., using the imaging conditions
listed in Tab. \ref{tab:Imaging-conditions}. The $C_{\mathrm{c}}$-correction,
notably, allowed to resolve the $\left\{ 2110\right\} $-family of
spatial frequencies not visible in a conventional $C_{\mathrm{s}}$
($C_{3})$-corrected electron microscope. A 20 minutes long time series
of holograms was acquired using a 2k by 2k CCD camera (Model 894 US1000,
Gatan Inc.), each with 8 seconds exposure time owing the great instrumental
stability of the microscope in order to enhance the SNR. We had to
slightly defocus the \emph{h}-BN sample plane to have sufficient contrast
for selecting the desired object position into the field of view.
We note that the defocus, as well as the two-fold astigmatism, were
considerably drifting and that the electron induced charging of the
sample is changing at a modest level over the time frame of the series.
Significant, presumably knock-on induced beam damage can be observed
over the 20 minutes, especially at the boundary to vacuum. The recorded
holograms were then processed off-line through a removal of dead and
hot pixels by an iterative local threshold algorithm, as well as a
masking out of Fresnel fringes \citep{Linck2009}. In addition, a
deconvolution of the CCD camera's MTF and a modest Wiener filtering
\citep{Linck2010a} in Fourier space were employed to increase the
SNR of the holograms \citep{Linck2010a}. Within the holographic Fourier
reconstruction method \citep{Lehmann2002}, one sideband was masked
with a circular tenth-order Butterworth filter with a radius of $8.5\,\mathrm{n}\mathrm{m}^{-1}$.
The phase of the reconstructed wave was subtracted by the phase reconstructed
from an additionally recorded and equally processed object-free empty
hologram, to correct for distortions induced by the fiber optics that
couples the scintillator to the CCD camera. 
\begin{table}
\begin{tabular}{|l|l|}
\hline 
acceleration voltage $U_{\mathrm{a}}$ &
80 kV\tabularnewline
\hline 
$C_{\mathrm{c}}$ and $C_{\mathrm{s}}(C_{3})$ &
$<10\,\text{\ensuremath{\mu}m}$\tabularnewline
\hline 
image spread $\sigma_{\mathrm{i}}$ &
40~pm\tabularnewline
\hline 
information limit &
0.13 nm\tabularnewline
\hline 
diffraction lens excitation &
65\%\tabularnewline
\hline 
pixel size of hologram &
0.054 nm\tabularnewline
\hline 
mean counts per hologram pixel $I_{\mathrm{hol}}$ &
$\sim10000$\tabularnewline
\hline 
biprism voltage $U_{\mathrm{bi}}$ &
160 V\tabularnewline
\hline 
fringe visibility $\mu\left(\boldsymbol{k}_{\mathrm{c}}\right)$\citep{Lubk2012} &
0.3\tabularnewline
\hline 
DQE$\left(\boldsymbol{k}_{\mathrm{c}}\right)$\citep{Lubk2012} &
0.5\tabularnewline
\hline 
\end{tabular}\caption{\label{tab:Imaging-conditions}Holographic imaging conditions at TEAM
I microscope adjusted for electron wave reconstruction of two-dimensional
materials.}
\end{table}
 The as-reconstructed amplitude and phase of a small region of overlapping
\emph{h}-BN sheets are depicted in Figs.~\ref{fig:Autocorrection-procedure.}a,b.
Since the sample is defocused, one observes an amplitude contrast
by the PCTF. Moreover, a rather large 2-fold astigmatism and other
residual aberrations seem to be present, rendering a quantitative
analysis almost impossible. We now apply the auto-correction procedure
outlined in Section~\ref{subsec:A-posteriori-correction}. Figs.~\ref{fig:Autocorrection-procedure.}c-j
show the results of the two auto-correction steps; the correction
of symmetric aberrations from the WPO property (Figs.~\ref{fig:Autocorrection-procedure.}e-g)
and the final correction including also antisymmetric aberrations
after exploiting the centrosymmetry of the \emph{h}-BN lattice (Figs.~\ref{fig:Autocorrection-procedure.}h-j).
Clearly, the amplitude (Fig.~\ref{fig:Autocorrection-procedure.}e)
is almost constant after removal of symmetric aberrations (Fig.~\ref{fig:Autocorrection-procedure.}a
and Fig.~\ref{fig:Autocorrection-procedure.}e are displayed within
the same greylevels), proving the experimental feasability of the
first auto-correction step. The numerical phase plate (Fig.~\ref{fig:Autocorrection-procedure.}c)
computed using Eq.~(\ref{eq:chi_s}) provides a good SNR only where
the reciprocal space is filled with specimen information (\emph{i.e.,
h}-BN systematic reflections, Fig.~\ref{fig:Autocorrection-procedure.}d).
These are, however, sufficient to determine the geometrical aberration
coefficients (within 95\% confidence intervals) of first-order aberrations,
namely defocus, $C_{1}=4.0\pm0.8\,\text{nm}$, and two-fold astigmatism,$\left\{ A_{1}=2.5\pm1.1\,\mathrm{nm},\alpha_{1}=54{^\circ}\pm25{^\circ}\right\} $,
by fitting a smooth polynomial 
\begin{equation}
\chi_{\mathrm{s}}=\frac{2\pi}{k_{0}}k^{2}(C_{1}+A_{1}\cos\left(2\alpha-\alpha_{1}\right))
\end{equation}
(Fig. \ref{fig:Autocorrection-procedure.}g) with the help of a Levenberg-Marquardt
algorithm (third order symmetric aberrations are small and could safely
be neglected). We note that the correction with the numerically obtained
phase plate (Fig. \ref{fig:Autocorrection-procedure.}c) yields almost
identical results as the correction with the corresponding fitted
phase plate (Fig. \ref{fig:Autocorrection-procedure.}g).

As stated above, the determination of the numerical antisymmetric
phase plate is merely possible for centrosymmetric specimen. Ignoring
the small difference in atomic species, \emph{h}-BN fulfills this
symmetry condition for two different points, the center of the BN
hexagons and the midpoints between the binding atoms. In order to
find the most centrosymmetric region of interest, the for symmetric
aberrations corrected phase was split up into sub images, containing
about 25 unit cells. Subsequently, for all of them, a numerical measure
for order the deviation from centrosymmetry was calculated. The most
centrosymmetric sub image was finally used to determine the antisymmetric
phase plate from Eq. (\ref{eq:chi_a}), from which the corresponding
phase plate (Fig. \ref{fig:Autocorrection-procedure.}j) 
\begin{align}
\chi_{a}=2\pi\frac{k^{3}}{k_{0}^{2}} & \left(\frac{1}{3}A_{2}\cos\left(3\alpha-\alpha_{A_{2}}\right)\right.\nonumber \\
 & \left.{\color{white}\frac{1}{1}}+B_{2}\cos\left(\alpha+\alpha_{B_{2}}\right)\right)
\end{align}
with a three-fold astigmatism of $\left\{ A_{2}=158\pm261\,\mathrm{nm},\alpha_{2}=-90\si{\degree}\pm16\si{\degree}\right\} $
and an axial coma of $\left\{ B_{2}=67\pm49\,\mathrm{nm},\alpha_{\mathrm{B_{2}}}=-20\si{\degree}\pm54\si{\degree}\right\} $
was fitted. Note that the 95\% confidence intervals computed from
the fit residual are rather large in this case, which is due to the
relatively small size of the symmetric patch used for the fitting
procedure.

\begin{figure}
\includegraphics[width=1\columnwidth]{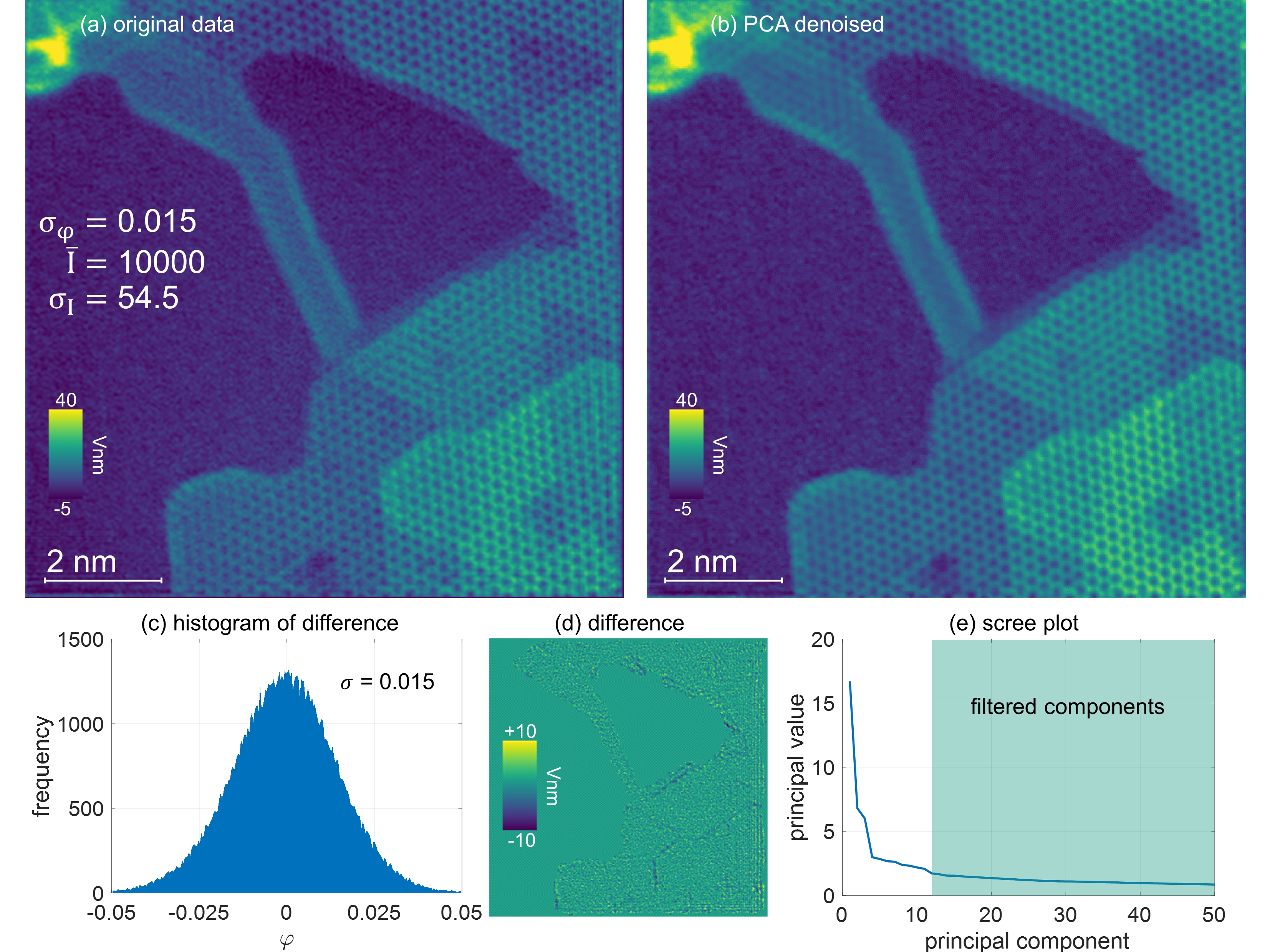}\caption{\label{fig:Denoising}Denoising of of the aberration-corrected \emph{h}-BN
phase image reconstructed by off-axis electron holography. (a) shows
original autocorrected data and corresponding standard deviations
of phase $\sigma_{\mathrm{\varphi}}$ and intensity $\sigma_{\mathrm{I}}$
in vacuum, (b) the PCA-denoised data after truncation to the first
11 principal components. The selection criterion is the last kink
in the scree plot (e). Both the difference (d) and the histogram (e)
reveal no noticeable deviation from Gaussian noise between the original
and denoised image except at some structurally fluctuating edges.}
\end{figure}

In order to prepare the autocorrected high-resolution phase data for
interpretation we apply PCA denoising in a final step (Fig. \ref{fig:Denoising}).
Again, no a-priori information about the material is required for
this procedure. We rather exploit the regular geometric structure
consisting of repeating honeycombs to create statistical data, which
can be treated by PCA (i.e., model-free) denoising \citep{Potapov2019}.
Our procedure consists of cutting out patches slightly larger than
one honeycomb, centering them and subjecting the stack of patches
to a PCA (see Appendix \ref{sec:PCA-Analysis} for details). Inspecting
the scree plot (Fig. \ref{fig:Denoising}(e)) we identified 11 non-noise
components and truncated the data accordingly. The thereby denoised
image is shown in Fig. \ref{fig:Denoising}(b) together with the original
data (Fig. \ref{fig:Denoising}(a)). The deviations to the original
data (Fig. \ref{fig:Denoising}(d)) exhibit a noise-like Gaussian
distribution (Fig. \ref{fig:Denoising}(c)) with the same standard
deviation as the original phase noise in vacuum. Moreover, it shows
no particular structure except at the edges, where fluctuations due
to radiation damage occur during acquisition. We finally note that
the the phase noise $\sigma_{\varphi}$ of the original data (Fig.
\ref{fig:Denoising}(a)) is consistent with that observed in the conventional
HRTEM image data $\sigma_{\mathrm{I}}$ (obtained from the noninterference
terms of the recorded hologram) after rescaling $\sigma_{\mathrm{\varphi}}^{2}\approx\text{DQE}\left(0\right)\text{DQE}\left(\boldsymbol{k}_{\mathrm{c}}\right)/\sigma_{\mathrm{I}}^{2}\mu_{\mathrm{c}}\left(\boldsymbol{k}_{\mathrm{c}}\right)^{2}$,
experimentally validating the noise considerations leading to Eq.
(\ref{eq:noise_ratio}). Here, the phase noise is smaller than the
noise in the ``raw'' phase image because of the slight Wiener filtering
mentioned above.

\section{Results and Discussion}

\begin{figure}
\vspace{-80bp}
\includegraphics[viewport=50bp 100bp 1010bp 794bp,width=1.35\columnwidth]{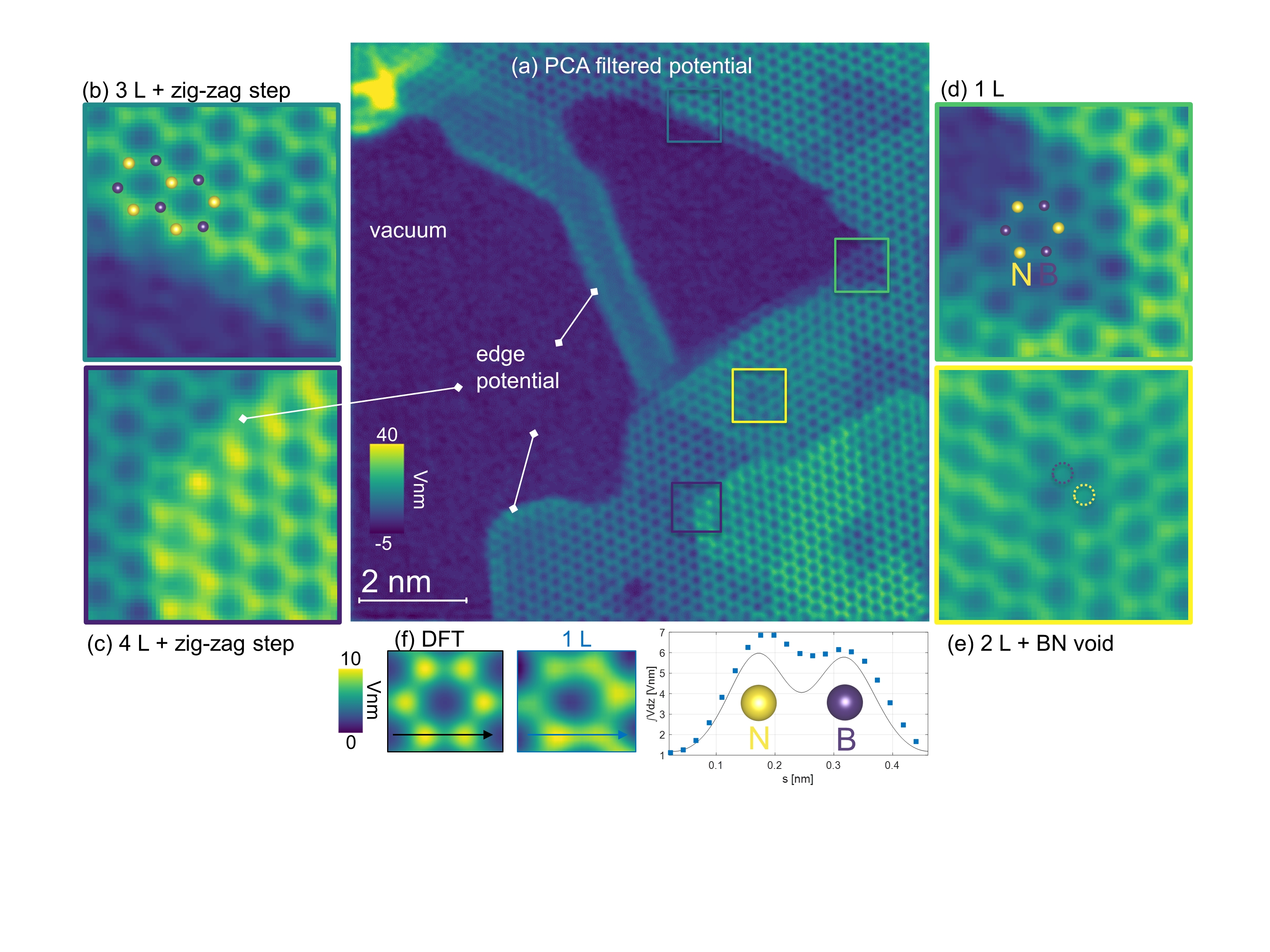}\caption{\label{fig:Potential-analysis.}High-resolution potential analysis
(a). Zoom-ins show zig-zag steps comprising two atomic layers (b,c),
a monolayer region (d) and a BN void defect (r). The monolayer potential
is depicted together with the DFT result in (f).}
\end{figure}

After successful application of the auto-correction and denoising
procedure we can now analyze the phase maps in detail. The high-resolution
data (Fig. \ref{fig:Potential-analysis.}) clearly reveals regions
of well-ordered honeycomb lattice with a multilayer morphology including
various defects and edge structures. There are also regions, notably
the bridge and some of the edges, where no or only a smeared-out honeycomb
lattice is visible (facets are discernible though). These coincide
with strongly oscillating parts of the sample, which were vanishing
during the time series due to the electron irradiation. We therefore
ascribe the loss of the high-spatial frequency data in this regions
to local vibrations rather than some sort of amorphization. Similar
observations and quantification of the lattice distortion at the edges
and vacancies have also been observed by quantitative phase contrast
imaging as well as STEM imaging \citep{Alem(2011),Alem2012}.

\begin{figure}
\vspace{-80bp}
\includegraphics[viewport=50bp 25bp 1058bp 794bp,width=1.45\columnwidth]{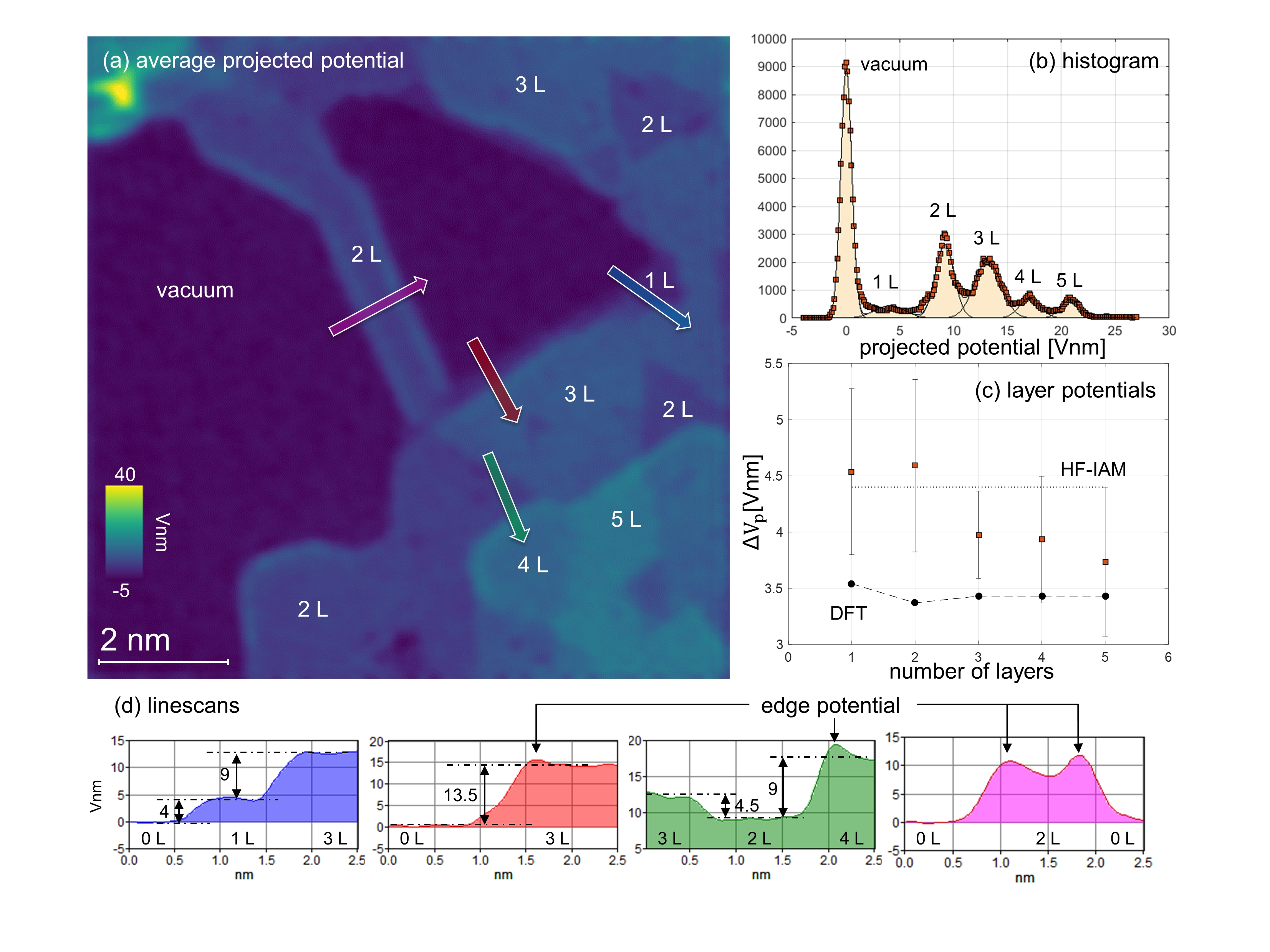}\caption{\label{fig:average-potential-analysis}Average potential analysis
of \textit{h}-BN. The overview image (b) contains several flakes of
different layer number, shaped by electron beam irradiation. The potential
data histogram (b) reveals the presence of several monolayers, seperated
by well-defined potential offsets $\Delta V$ (c). The linescanes
(d) provide local measures for the latter and reveal the presence
of potential bumps at edges and steps.}
\end{figure}

Indeed, the quantitative phase data allows for a direct comparison
with the projected potential of the \textit{h}-BN lattice. We first
focus on the low frequency information (which might have been obtained
also with medium resolution holography modes and without the autocorrection
of aberrations). To that end the high-resolution data is convoluted
with a round top-hat function of the radius of a lattice constant
(0.25 nm) and divided by $C_{\mathrm{E}}$ (see Eq. (\ref{eq:PGA}))
yielding the averaged projected Coulomb potential corresponding to
the zeroth Fourier component of the potential in a periodic lattice.
It depends sensitively on the charge (de)localization due to chemical
bonding \citep{OKeeffe1994} (see also Appendix \ref{sec:Mean-Inner-Potential})
and is proportional to the diamagnetic susceptibility according to
the Langevin theory\citep{Ibers1958} amongst others. In our case
the average potential data reveals (Fig. \ref{fig:average-potential-analysis}(a))
the presence of different sample thicknesses, ranging from one to
five atomic layers,  visible as areas of constant projected potential,
interrupted by a small number of defects. In the histogram Fig. \ref{fig:average-potential-analysis}(b)
these regions can be associated to clear distinct peaks. Moreover,
the differences of the peak positions determined by Gaussians fits
yield the average potentials of each layer (Fig. \ref{fig:average-potential-analysis}(c)),
that show a small decrease towards higher layer numbers. Noting that
the average potential corresponds approximately to the sum of the
second spatial moment (i.e., spatial extension, see Appendix \ref{sec:Mean-Inner-Potential}
for a derivation) of the charge distribution of the contributing atoms,
i.e., $\bar{V}\sim\sum\left\langle r^{2}\right\rangle $$_{\mathrm{at}}$,
this decrease in average potential could reflect a growing localization
of the out-of-plane orbitals ($3p_{\mathrm{z}}$ orbitals) in between
\textit{h}-BN layers as compared to the free surfaces. Indeed, the
mono- and bilayer average projected potential of \textasciitilde$4.5\,\mathrm{Vnm}${}
rather fit with independent atom potentials computed from Hartree-Fock
\citep{Weickenmeier1991} ($4.4\,\mathrm{Vnm}$). The latter tend
to be too delocalized compared to those computed from a full density
functional theory (DFT) calculation (using FPLO-18\citep{Koepernik(1999)},
see Appendix \ref{sec:DFT}) including chemical bonds and correlation  that
agrees with a value of $3.4\,\mathrm{Vnm}$ better to three and more
layers(Fig. \ref{fig:average-potential-analysis}(c)). Note furthermore
that the structurally and atom-weigth-wise closely related graphene
has a projected potential of 4.5 Vnm, which also reflects the stronger
delocalization of the shell electrons in conducting graphene. Another
possible explanation of the large potential values could be the positive
charging of the (insulating) \textit{h}-BN in the beam (ejection of
secondary electrons).

A second noticeable feature is the potential increase visible at the
edges and steps of the sample (most prominent in the doublelayer bridge
region). Different physical effects may be attributed to this potential
elevation:
\begin{enumerate}
\item Formation of (covalent) interlayer bonds at the zig-zag edges of bilayer
\textit{h}-BN, as proposed by Alem \textit{et al. }\citep{Alem2012},
through the following mechanisms: a local compression of the projected
atomic positions or the tilting of the covalent B-N bonds out of plane,
both yield a raised projected potential. These interlayer bonds could
also explain the enhanced stability of even numbered layers under
electron irradiation and thus their dominant appearance in the data
set.
\item Delocalization of in-plane $s,\,p_{\mathrm{x}}\,,p_{\mathrm{y}}$
orbitals into vacuum, which could potentially lead to the observed
increase of the average potential. Indeed, DFT calculations reported
in literature predict the emergence of metallic edge states at the
zig-zag edges \citep{Barone2008,Zeng2010}.
\item Systematic adhesion of residual gas atoms with different atomic potentials
at the edges, such as Oxygen \citep{Lopez-Bezanilla2011}.
\end{enumerate}
Further studies are necessary to clarify and disentangle those effects
quantitatively. 

We now turn to the analysis of the autocorrected and denoised high-resolution
data shown in Fig. \ref{fig:Potential-analysis.}(a). The denoised
high-resolution data of the monolayer allows to distinguish between
the B and N sites in the monolayer (Fig. \ref{fig:Potential-analysis.}(d)).
Comparing the holographically measured potentials with the ab-initio
potentials, which have been smeared out by multiplying the envelope
function pertaining to the TEAM I instrument at 80 kV (see Fig. \ref{fig:transfer-functions}(b)),
we observe good agreement with an additional smearing of the experimental
data along the bonding directions (Fig. \ref{fig:Potential-analysis.}(f)).
Whether this is due to thermal vibrations (not included in our analysis)
or other damping factors remains an open question at this stage. Turning
to the edge structures, we observe that the zig-zag boundary is the
prevalent configuration (Fig. \ref{fig:Potential-analysis.}(b),(c)),
which coincides with the ab-initio predictions \citep{Fu2017}. We
further note occasional distortions of the edge lattice, which may
be attributed to some out-of-plane bending or ongoing beam damage
(c.f. \citep{Alem(2011),Alem2012}). Moreover, almost all edges reveal
an increase in projected potentials, which has been already discussed
above . That notably also includes steps (e.g., the step from 2 L
to 4 L in Fig.\ref{fig:Potential-analysis.}(c)). As noted previously
we attribute these localized edge potentials to an increased electron
delocalization, most probably due to a reconstruction of the edge
structure along $z$-direction including the formation of interlayer
covalent bonds\textit{ }\citep{Alem2012}. A detailed comparison to
the emergence of particular edge states and the electronic configuration
of defects (e.g., BN void depicted in Fig. \ref{fig:Potential-analysis.}(e))
is, however, beyond the scope of this work and will be conducted elsewhere.

Summing up we showed, how autocorrected off-axis holography may be
used as a high-resolution and dose efficient probe for 2DMs. In this
regard we performed the first parameter-free correction for antisymmetric
aberrations based on the afore-mentioned symmetry principles. Notably
the presented aberration autocorrection scheme and the favorable noise
transfer properties of EH facilitates the reconstruction of quantitative
potential data over a large spatial frequency band extending from
zero to atomic resolution. The autocorrection scheme facilitates a
removal of residual aberrations and defocus without the need to separately
measuring or fine tuning them, rendering it a suitable method for
\textit{in-situ} studies, where long time-series need to be recorded.
The high-structural order of 2DMs furthermore facilitated an efficient
suppression of noise by adopting a PCA noise removal algorithm. Using
these capabilities, we reveal several properties of \textit{h}-BN.
Notably, the electronic orbitals in \textit{h}-BN are significantly
more localized as in the structurally similar graphene, resulting
in a comparatively low mean projected potential of 3.6 Vnm. We could
confirm that edges favor the zig-zag configuration and found a peculiar
localized increase of the potential at the edges. The latter is attributed
to the delocalization of electron edge states.

\section*{Acknowledgements}

We are grateful to Tore Niermann, who supported us at recoring additional
hologramms of another \textit{h-}BN sample. We acknowledge funding
from the European Research Council (ERC) under the Horizon 2020 research
and innovation program of the European Union (grant agreement number
715620) and the Deutsche Forschungsgemeinschaft (DFG, German Research
Foundati- on) -- Project-ID 417590517 -- SFB 1415. This work is
also supported by the Office of Science, Office of Basic Energy Sciences
of the U.S. Department of Energy under Contract No. DE-AC02---05CH11231. This work was kindly supported by the Initiative and Networking Fund of the Helmholtz Association of German Research Centers through the International Helmholtz Research School for Nanoelectronic Networks, IHRS NANONET (VH-KO-606).

 \begin{appendix}
\section{WPO Phase Plate\label{sec:WPO-Phase-Plate}}

This appendix contains derivations for the expressions (\ref{eq:chi_s})
and (\ref{eq:chi_a}) relating holographic data and aberrations. We
start with rewriting the expression for the suitably normalized Fourier
transformed image wave (\ref{eq:img_Wave_k}) using $\tilde{\varphi}=A_{\tilde{\varphi}}e^{iP_{\tilde{\varphi}}}$

\begin{align}
\frac{\tilde{\varPsi}_{\mathrm{img}}-2\pi\delta\left(\boldsymbol{k}\right)}{A} & =\tilde{A}_{\mathrm{img}}e^{i\tilde{\phi}_{\mathrm{img}}}\\
 & =iA_{\mathrm{\tilde{\varphi}}}\left(\boldsymbol{k}\right)e^{-i\left(\chi_{\mathrm{s}}\left(\boldsymbol{k}\right)+\chi_{\mathrm{a}}\left(\boldsymbol{k}\right)-P_{\tilde{\mathrm{\varphi}}}\left(\boldsymbol{k}\right)\right)}E\left(\boldsymbol{k}\right)\,.\nonumber 
\end{align}
Since $\varphi\left(\boldsymbol{r}\right)$ is a real function we
have $P_{\tilde{\varphi}}\left(\boldsymbol{k}\right)=-P_{\tilde{\varphi}}\left(-\boldsymbol{k}\right)$
and hence
\begin{equation}
P_{\tilde{\varphi}}\left(\boldsymbol{k}\right)\mathrm{mod}2\pi+P_{\tilde{\varphi}}\left(-\boldsymbol{k}\right)\mathrm{mod}2\pi=2\pi n,\,n\epsilon\mathbb{N}.
\end{equation}
Consequently,
\begin{equation}
\chi_{\mathrm{s}}(\boldsymbol{\boldsymbol{k}})=-\frac{1}{2}\left(\tilde{\phi}_{\mathrm{img}}\left(\boldsymbol{\boldsymbol{k}}\right)+\tilde{\phi}_{\mathrm{img}}\left(-\boldsymbol{\boldsymbol{k}}\right)\right)+\frac{\pi}{2}+\pi n\,.\label{eq:chi_s2}
\end{equation}
Following a similar line of reasoning and assuming an centrosymmetric
object, i.e., $P_{\tilde{\varphi}}\left(\boldsymbol{k}\right)\epsilon\left\{ 0,\pi\right\} $
we have
\begin{equation}
\chi_{\mathrm{a}}(\boldsymbol{\boldsymbol{k}})=-\frac{1}{2}\left(\tilde{\phi}_{\mathrm{img}}\left(\boldsymbol{\boldsymbol{k}}\right)-\tilde{\phi}_{\mathrm{img}}\left(-\boldsymbol{\boldsymbol{k}}\right)\right)+\pi n
\end{equation}
Here the $\pi n$ stems from the possible $\pi$ phases of the real
centrosymmetric object.

\section{Noise Transfer\label{sec:Noise-Transfer}}

In the derivation of the SNR for off-axis holography and conventional
phase contrast HRTEM we used the generalized Lenz model (uncorrelated
shot and detector noise, commensurable sinc sideband mask, noise small
compared to total intensity), which gives good agreement with the
observed phase noise, in particular under weak contrast conditions
as present in the WPO \citep{Roder(2014)a}. Moreover, we assumed
that the noise characteristics (e.g., variance) do not depend significantly
on the position on the detector, which is again a good approximation
for weakly scattering objects. Using these approximations the variance
of the reconstructed phase reads
\begin{equation}
\sigma_{\varphi}^{2}=\frac{1}{I\mu_{\mathrm{c}}\left(\boldsymbol{k}_{\mathrm{c}}\right)^{2}\text{DQE}\left(\boldsymbol{k}_{\mathrm{c}}\right)}
\end{equation}
Here, the DQE denotes a 2D generalization of detection quantum efficiency
defined as 
\begin{equation}
\text{DQE}\left(\boldsymbol{k}_{\mathrm{c}}\right)=\frac{\text{MTF}^{2}\left(\boldsymbol{k}_{\mathrm{c}}\right)}{\text{NPS}\left(\boldsymbol{k}_{\mathrm{c}}\right)}\,,
\end{equation}
with the MTF denoting the modulation transfer and NPS the (normalized)
white noise power spectrum of the detector. The phase SNR then reads
\begin{equation}
\mathrm{SNR_{hol}}=\frac{\varphi_{\mathrm{obj}}}{\sigma_{\varphi}}=\sqrt{\frac{E^{2}\varphi_{\mathrm{obj}}^{2}I\mu_{\mathrm{c}}\left(\boldsymbol{k}_{\mathrm{c}}\right)^{2}}{\text{DQE}\left(\boldsymbol{k}_{\mathrm{c}}\right)}}\,.
\end{equation}
The noise analysis for conventional weak phase contrast HRTEM starts
with the shot noise amplified by the detector
\begin{equation}
\sigma_{\mathrm{I}}^{2}=I\text{NPS}\left(0\right)
\end{equation}
from which the SNR is readily derived inserting the relation between
$I$ and the phase (PCTF)
\begin{equation}
\mathrm{SNR_{conv}}=\sqrt{\frac{I^{2}\mathrm{PCTF}^{2}\varphi_{\mathrm{obj}}^{2}}{I\text{DQE}\left(0\right)}}\,.
\end{equation}
We finally arrive at for the ratio between both as noted in the main
text
\begin{equation}
\frac{\mathrm{SNR_{hol}}}{\mathrm{SNR_{conv}}}=\frac{\mu_{\mathrm{c}}\left(\boldsymbol{k}_{\mathrm{c}}\right)\sqrt{\text{DQE}\left(0\right)}}{\sin\chi_{\mathrm{s}}\sqrt{\text{DQE}\left(\boldsymbol{k}_{\mathrm{c}}\right)}}\,.
\end{equation}

\section{Mean Inner Potential\label{sec:Mean-Inner-Potential}}

Establishing a well-defined relationship between averaged projected
potential or projected mean inner potential of a 2DM and the charge
distribution holds some pitfalls in the infinite crystal limit \citep{Kleinman1981}.
Indeed, a mean inner potential is not well-defined in this case and
depends on fixing boundary conditions or a reference. To circumvent
this problem, ab-inito calculations of MIPs of bulk crystals have
been carried out for slab geometries, containing a sufficiently large
vacuum region fixing the reference. In case of a finite crystal (as
observed experimentally), we may start off with dividing the 2DM domain
into ``atomic'' cells,
\begin{equation}
V\left(\boldsymbol{r}\right)=\sum V_{\mathrm{at}}\left(\boldsymbol{r}\right)\,,
\end{equation}
which shall contain one atom each but are not further specified at
this stage. The projected average of the potential over a certain
area $A$ , i.e.
\begin{equation}
\frac{1}{A}\int_{A}V\left(\boldsymbol{r}\right)d^{3}r=\frac{1}{A}\int\sum V_{\mathrm{at}}\left(\boldsymbol{r}\right)d^{3}r
\end{equation}
can now be computed as a sum of the atomic contributions, which are
contained within the area. This is most conveniently done in Fourier
space employing the Poisson equation
\begin{equation}
k^{2}\tilde{V}_{\mathrm{at}}\left(\boldsymbol{k}\right)=\frac{\tilde{\rho}_{\mathrm{at}}\left(\boldsymbol{k}\right)}{\varepsilon_{0}}\,.
\end{equation}
The atomic averaged potential now corresponds to the value at zero
spatial frequency
\begin{equation}
\tilde{V}_{\mathrm{at}}\left(k=0\right)=\underset{k\rightarrow0}{\lim}\frac{\tilde{\rho}_{\mathrm{at}}\left(k,\varphi_{k},\theta_{k}\right)}{\varepsilon_{0}k^{2}}\,,
\end{equation}
which is not determined in the required limit as both nominator and
denominator tend to zero. To solve that expression we may apply l'Hospital's
rule to the average over the full solid angle of the previous expression
(to remove the $\varphi,\theta$ dependency)
\begin{align}
 & \tilde{V}_{\mathrm{at}}\left(k=0\right)=\\
 & \underset{k\rightarrow0}{\lim}\partial_{k}^{2}\frac{\left\langle \tilde{\rho}_{\mathrm{at}}\left(k\right)\right\rangle _{\varphi_{k},\theta_{k}}}{2\varepsilon_{0}}\nonumber \\
= & \underset{k\rightarrow0}{\lim}\frac{1}{2\varepsilon_{0}}\partial_{k}^{2}\int d^{3}r\rho_{\mathrm{at}}\left(\boldsymbol{r}\right)\nonumber \\
 & \left\langle e^{ikr(\sin\theta_{k}\sin\theta_{r}\cos(\varphi_{k}\text{\textminus}\varphi_{r})+\cos\theta_{k}\cos\theta_{r})}\right\rangle \nonumber \\
= & -\frac{1}{2\varepsilon_{0}}\int dr^{3}r^{2}\rho_{\mathrm{at}}\left(\boldsymbol{r}\right)\nonumber \\
 & \left\langle (\sin\theta_{k}\sin\theta_{r}\cos(\varphi_{k}\text{\textminus}\varphi_{r})+\cos\theta_{k}\cos\theta_{r})^{2}\right\rangle \nonumber \\
= & -\frac{1}{2\varepsilon_{0}}\int dr^{3}r^{2}\rho_{\mathrm{at}}\left(\boldsymbol{r}\right)\nonumber \\
 & \left\langle \sin^{2}\theta_{k}\sin^{2}\theta_{r}\cos^{2}(\varphi_{k}\text{\textminus}\varphi_{r})+\cos^{2}\theta_{k}\cos^{2}\theta_{r}\right\rangle \nonumber \\
= & -\frac{2\pi}{3\varepsilon_{0}}\int drr^{2}\rho_{\mathrm{at}}\left(\boldsymbol{r}\right)\nonumber 
\end{align}
Note that we had to apply l'Hospital's rule twice, which requires
both the zeroth and any first moment with respect to $r$ to vanish.
Consequently, the ``atomic'' cells, used for partitioning the 2D
domain, have to be charge neutral and dipole-moment-free. That condition
restricts the allowed shapes and positions of the initially undefined
atomic patch choice. Noting that dipole-free atomic cells are centered
closely to the atomic positions in \textit{h}-BN, we have that the
averaged potential is a measure for the size of the electron cloud
around the atoms.

\section{DFT\label{sec:DFT}}

Density-functional \citep{Hohenberg(1964)} band-structure calculations
using the all-electron full-potential local-orbital (FPLO-18) \citep{Koepernik(1999)}
calculation scheme were employed to obtain the electronic properties
(e.g., electron density and potential) of single- to fivelayer \textit{h}-BN
and Graphene for reference. The calculations were scalar relativistic
\citep{Eschrig(2004)} and used the generalized gradient approximation
(GGA) of the exchange-correlation functional due to Perdew-Burke-Enzerhof
\citep{Perdew(1996)}. The in-plane structural parameter $a=2.505$ \AA \,
 and the distance between the layers $d=3.324$  \AA \, of \textit{h}-BN,
where taken from literature \citep{Paszkowicz2002} (and agree well
with our experimental findings).

\section{PCA Analysis\label{sec:PCA-Analysis}}

Principal component analysis (PCA) allows to find the optimal (w.r.t.
the Euclidean distance) linear decomposition of a statistically varying
signal into a truncated basis. It is a well-established method in
advanced statistical analysis and machine learning\citep{Jolliffe2016},
provided that a sufficiently large statistical set of signals can
be collected. In TEM it finds application in the analysis of EELS
and EDX data amongst others\citep{Titchmarsh1996,Parish2010,Potapov2019}.
Here we apply it to the autocorrected phase dataset decomposed into
patches containing one honeycomb of the 2DM structure. Due to the
high structural order of 2DMs these patches contain a finite number
of different species, namely 1-4 layer ``bulk'' honeycombs and corresponding
edges/steps. Because several thousand honeycombs are contained in
one hologram, the patches form a sufficiently large statistical basis,
except some particular edge states, which were too sparse to come
out in the PCA. In the PCA analysis we closely followed the steps
layed out in Ref. \citep{Potapov2019}:
\begin{enumerate}
\item Identification of patches by locating the honeycomb minima. In an
iterative procedure the patch origin is refined by aligning them with
their Center of Mass. We also subtracted the average of each patch
(which amounts to removing the first principal component).
\item PCA of the data matrix, whose rows correspond to the different patches
and the columns represent the interlaced spatial coordinates of the
patches.
\item Truncation of the decomposition by removing all components beyond
a kink in the scree plot (showing the magnitude-ordered principal
components). This filter is referred to as truncated PCA in literature.
\item Computation of the difference between original and truncated data,
confirming the Gaussian noise nature of the remainder (see Fig. \ref{fig:Denoising}(c)
in the main text).
\item Replacement of the patches in the original image with the PCA truncated
patches (Fig. \ref{fig:Denoising}(b))
\end{enumerate}
\end{appendix}

\section*{Bibliography}

\bibliographystyle{apsrev4-1}
\bibliography{2D}
\end{document}